\documentstyle[aps]{revtex}

\begin{document}
\input epsf
\draft
\twocolumn[\hsize\textwidth\columnwidth\hsize\csname
@twocolumnfalse\endcsname
\preprint{PURD-TH-98-10}
\date{September 28, 1998}
\title{Matter Creation via Vacuum Fluctuations
in the Early Universe and Observed  Ultra-High Energy
Cosmic Ray Events
}

\author{Vadim Kuzmin$^{a,}$\footnote{email address:
kuzmin@ms2.inr.ac.ru}  and Igor Tkachev$^{a,b,}$\footnote{ email address:
tkachev@physics.purdue.edu }}
\address{
$^a$Institute for Nuclear Research, Russian Academy of Sciences,\\
60th October Anniversary Prosp. 7a, Moscow 117312, RUSSIA \\
$^b$Department of Physics, Purdue University, West Lafayette, IN 47907, USA
}

\maketitle

\begin{abstract}
Cosmic rays of the highest energy, above the  Greisen--Zatsepin--Kuzmin
cut-off of the spectrum,  may originate in decays of superheavy
long-living X-particles. These particles may be
produced in the early Universe from vacuum fluctuations
during or after inflation and may  constitute a considerable
fraction of Cold Dark Matter. We calculate numerically their abundance
for a wide range of models.
X-particles are considered to be either bosons or fermions.
Particles that are several times heavier than inflaton,
$m_{\rm inflaton} \approx 10^{13}$ GeV, and were produced
by this mechanism, can account for the critical mass
in the Universe naturally. In some cases
induced isocurvature density fluctuations can leave an imprint in 
anisotropy of cosmic microwave background radiation.
\end{abstract}


\vskip2pc]


\section{Introduction}
According to the Greisen-Zatsepin-Kuzmin \cite{gzk} (GZK) observation,
the energy spectrum of Ultra High Energy (UHE) cosmic rays produced at
extragalactic distances should exhibit an
exponential cut-off at energy $E \sim 5 \times 10^{10}$ GeV.
However, a number of cosmic ray events with energies well beyond the
predicted GZK cut-off were recently observed by the various experimental
groups \cite{cr}.
This is an obvious contradiction with the standard cosmological
and particle physics models and clearly requires some new physics
beyond the Standard Model.

Conceptually, the simplest explanation \cite{KR,BKV} could be that
the highest energy cosmic rays are produced in the decays
of heavy long-living particles in cosmologically local
part of the Universe. We will call these
progenitors of the cosmic rays as X-particles. The mass of
$X$-particles has to be very large, $m_{X} \agt 10^{13}$~GeV,
for them to be responsible for the cosmic rays events in the energy range
$E \agt 10^{11}$ GeV.
It was noticed \cite{KT98,CKR} that such heavy particles are produced
in the early Universe from the vacuum fluctuations and their
abundance can be correct naturally, if the standard Friedmann epoch in
the Universe evolution was preceded by the inflationary stage.
Basically, it is the same process which during inflation had generated
primordial large scale density perturbations and seeded formation
of galaxies and galaxy clusters. However, there is a difference.
Inflationary stage is not required to produce superheavy particles
from the vacuum,
unlike to the case of generation of the long wavelength perturbations.
Rather, the inflation provides a cut off in excessive production of heavy
particles which would happen in the Friedmann Universe if it would
start from
the initial singularity \cite{KT98}. Note that we consider sterile or
almost sterile X-particles, so that their production in usual plasma
interactions and decays can be neglected. Thermal production of heavy
X-particles was discussed in Refs. \cite{KR,BKV} and production
during reheating considered recently in \cite{CKR2}.

Particle production from vacuum fluctuations in the Friedmann
Universe during
matter or radiation dominated stages had been considered long ago
\cite{pc_pwl,MMF} and this gives the basis for simple estimates
of X-particle abundance \cite{KT98} which have the mass
of oder or smaller than the Hubble constant at the end of inflation,
$m_X \alt H_i$. However, the most interesting case of
heavier X-particles, $m_X > H_i$, requires detailed calculations.
Relevant calculations were done already by
Chung, Kolb and Riotto \cite{CKR}.
However, to describe the end of inflation and the transition to the
Friedmann
universe, they match the fixed de Sitter background to the subsequent
radiation or matter dominated expansion either as an instantaneous
junction or with the help of a smoothing function. The result
is especially sensitive to the details of  the junction procedure in
the case of large $m_X$ which is of interest for us here. Further,
the range of
models considered in Ref. \cite{CKR} was restricted to the case
of scalar particles with conformal coupling to gravity.

The purpose of the present paper is to calculate production of
superheavy particles from the vacuum in the inflationary Universe
for a wide range of models, and to give detailed and extended
discussion of our other results already reported in letter \cite{KT98}.
Considering particle creation, we do not make any
approximations. To this end we find numerically the
exact evolution of the scale factor in the model of the ``chaotic''
inflation \cite{al83}.
Our basic formalism relies on the method of Bogolyubov transformations,
which for the case of particle creation by non-stationary gravitational
field was developed in refs. \cite{pc,pc_pwl}. We consider
the possibility for X-particles to be either bosons
or fermions.

In Section II we remind the reader about the  explicit relations between
mode functions of the field, coefficients of the Bogolyubov transformation
and the particle number. In Section III we present results of our numerical
calculations of particle production from the vacuum fluctuations both
for Friedmann and inflationary stages in the universe evolution.
In Section IV we relate results of previous section to the present-day
density of X-particles. In Section V we discuss relevance of our
results to the UHE cosmic rays events, and Section VI contains our
conclusions.

\section{Particle production by a non-stationary gravitational field}

Here we summarize the basic formalism of gravitational
particle creation in an
expanding universe which we employed in our analysis.
For more details see Refs. \cite{pc_pwl,pc}.
We choose the metric to be conformally flat at an early cosmological
epoch,
$ds^2 = a(\eta )^2 (d\eta^2 - d{\bf x}^2)$.
The number density of $X$-particles created from the vacuum in a
time varying
cosmological background can be written as
\begin{equation}
n_X=\frac{1}{2\pi^2a^3} \sum_s\int |\beta_k|^2 k^2 dk \, \, ,
\label{nX}
\end{equation}
where $\beta_k$ are the Bogolyubov's coefficients which
relate ``in'' and ``out'' mode functions, $k$ is the comoving momentum,
and $\sum_s$ is the sum over spin states. The expression (\ref{nX})
gives the number
density of particles only, with the equal amount of antiparticles being
created in the case of charged fields. The creation of Bose and Fermi
particles is to be considered separately.

i) {\it Bosons.}
The  mode functions, $\chi_{k} \equiv \chi_{k}(\eta )$
of a scalar Bose field are solutions of the oscillator equation
\begin{equation}
{\chi ''}_{k} + \omega^2_k \chi_{k} = 0 \,\, ,
\label{lin}
\end{equation}
with the time dependent frequency
\begin{equation}
\omega^2_k = k^2 - \frac{a''}{a} (1-6\xi) + m_X^2 a^2
\,\, .
\label{ome}
\end{equation}
Here $' \equiv d/d\eta$. The constant $\xi$ describes direct coupling
to the curvature, with $\xi=0$ corresponding to the minimal coupling
and $\xi =1/6$ being the case of conformal coupling.
Equations for
massless conformally coupled quanta are reduced to the equations
in flat space-time and such particles are not created.
For massive particles conformal invariance is broken and particles
are created regardless of value of $\xi$.
Given the initial (vacuum) conditions
\begin{equation}
{\chi }_{k}(0) = \omega^{-1/2},\; \; \;
{\chi '}_{k}(0)=-i\omega \chi_{k} \, ,
\label{incon}
\end{equation}
the Bogolyubov's coefficients at any time moment $\eta$ are found to be
\begin{equation}
|\beta_k|^2 = \frac{ |{\chi '}_{k}|^2 + \omega^2 |\chi_{k}|^2
-2 \omega } {4\omega} \, .
\label{beta}
\end{equation}

ii) {\it Fermions.}
The relevant mode functions of the Fermi field satisfy the oscillator
equation with the complex frequency
\begin{equation}
{\chi ''}_{k} + (\omega^2_k - i m_X a')\chi_{k}= 0 \,\, ,
\label{lin_f}
\end{equation}
where the real part of the frequency is given by
$\omega^2_k = k^2 + m_X^2 a^2 $.
We choose
\begin{equation}
{\chi }_{k} (0) = \sqrt{1 - \frac{m_Xa}{\omega} },\; \; \;
{\chi '}_{k} (0) =-i\omega \chi_{k} \, ,
\label{inconf}
\end{equation}
as the initial conditions. In this case we find per spin state
\begin{equation}
|\beta_k|^2 =
\frac{\omega -m_X a - {\rm Im}(\chi_{k} \chi^{*'}_{k})}{2\omega} \, .
\label{beta_f}
\end{equation}
Summation over spins gives the factor of 2 in Eq. (\ref{nX}).

\section{Numerical results}

\subsection{Power-Law Cosmology}

In this section we consider particle creation during the regular, i.e.
non-inflationary, stage of the Universe expansion. We shall refer
to this stage as the Friedmann stage of the Universe evolution.
During the Friedmann stage the scale factor is given by the expression
$a(t) \propto t^\alpha$ with $\alpha < 1$. This case is of interest since
$\alpha=1/2$ and $\alpha=2/3$ correspond to the radiation and
matter dominated stages of the Universe expansion, which are inevitable.
Moreover, in some cases (and we shall encounter such situations below)
particle creation during this stage gives a dominant
contribution, while creation during inflation (or at the
transition between stages) is negligible.

\begin{figure}
\leavevmode\epsfysize=5.5cm \epsfbox{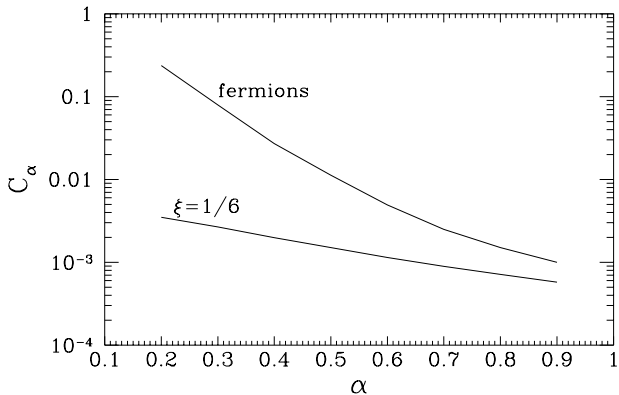}
\caption{Coefficient $C_\alpha$, defined in Eq. (9),
is shown as a function of $\alpha$ for background cosmology
with a power law scale factor $a \propto t^\alpha$.
}
\label{fig:C}
\end{figure}

Let us first consider massive particles conformally coupled to gravity.
It is the particle mass which couples the system to the
background expansion and serves as the source of particle
creation. Therefore we expect $n_X \propto m_X^3 a^{-3}$ at late
times when
particle creation diminishes and the number of particles
conserves in a
comoving volume. Because the scale factor has an arbitrary
normalization, for
definiteness it is
more convenient to rewrite this expression using relations
$a \propto (mt)^\alpha$, or $a \propto (m/H)^\alpha$. We prefer the
latter  choice since the exact value of time depends on the previous
history
in the inflationary cosmology and consequently the time coordinate,
$t$, is
ambiguously defined. Moreover, it is a non-zero value of the
Hubble constant which is responsible for particle creation
in an expanding universe.
The particle production is expected to stop when
$H \ll m_X$.
Therefore, at the late epoch we can parameterize the anticipated formulae
for $n_X$ as
\begin{equation}
n_X = C_\alpha m_X^3 \left(\frac{H}{m_X} \right)^{3\alpha} \, ,
\label{n_fr}
\end{equation}
where the constant $C_\alpha$ depends only upon the background cosmology,
i.e. $\alpha$, and it can be found numerically. Note that for the
radiation
dominated case it was found in Ref. \cite{pc_pwl},
$n_X \approx 5.3 \times 10^{-4} m_X^3 \, (m_X t)^{-3/2}$.
We confirm this result and extend it here to a broad range of
cosmologically
interesting values of $\alpha$, which includes the matter dominated case
as well.
Our results are summarized in Fig~\ref{fig:C} where we plot $C_\alpha$
as a function of $\alpha$.
The particle number reaches this asymptotic value, Eq. (\ref{n_fr}),
if the initial vacuum
state was defined at the sufficiently early epoch, $H(0) \gg m_X$.
We found that $H(0) \approx 10^4 m_X$ is sufficient for the results
to become
practically independent of the chose of the initial time
if $\alpha =1/2$, with
smaller (larger) ratio of $H(0)/m_X$ required at larger (smaller)
$\alpha$.

For the radiation dominated Universe one finds $\Omega_X \equiv
\rho_X/\rho_c = m_X n_X \, 32\pi Gt^2/3$ with the present
value of $\Omega_X$ being equal to
$\Omega_X \sim 2\times 10^{-2} (m_X^2/M^2_{\rm Pl})\sqrt{m_X t_e}$,
where $t_e$ is the time of equal densities of radiation
and matter in the $\Omega =1$ Universe. This gives
$\Omega_X \sim (m_X/10^9 {\rm GeV})^{5/2}$. We see that stable
particles with $m_X \agt 10^9$ GeV
will overclose the Universe even if initially they were in a vacuum state
and were created from the vacuum during
the regular Friedmann radiation dominated stage of the evolution.
(It is possible to separate the vacuum creation from the creation in
collisions
in plasma since X-particles may be effectively sterile.)

However, this restriction will be not valid\footnote{Another possibility
is the late time entropy release.} if the evolution of the Universe,
as it is believed, was more complicated than the simple radiation
dominated
expansion from a singularity. The Hubble constant may have
never exceeded $m_X$, which is the case of inflation,
$H(0) \approx m_\phi \approx 10^{13}$ GeV.
Moreover, compared to the case considered above,
the density of X-particles created during inflation is additionally
diluted
by the entropy release in reheating after inflation.

\subsection{Inflationary Cosmology}

Any viable modern cosmological model invokes the hypothesis of
inflation (for a review and list of references, see refs.
\cite{inflation,KT_book}). During inflation the Universe expands
exponentially
which solves the horizon and flatness problems of standard Big-Bang
cosmology. Inflation is generally assumed to be driven by the special
scalar field $\phi$ known as the {\it inflaton}.   Fluctuations
generated at
inflationary stage may have the strength and the power spectrum
suitable for
generation of the large-cale structure. This fixes the range of
parameters of the inflaton potential. For example, the mass
of the inflaton field has to be $m_\phi \sim 10^{13}$ GeV. During
inflation,
the inflaton field
slowly rolls down towards the minimum of its potential. Inflation
ends when
the potential energy associated with the inflaton field becomes
smaller than
the kinetic energy. Coherent oscillations of the inflaton field contained
all the energy of the Universe at that time. It is possible \cite{KLS}
that a significant fraction of this energy was
released to other boson species after only a dozen oscillations of the
inflaton field, in the regime of a broad parametric resonance.
This process was studied in detail \cite{KT,KLS2}. It was shown that
even rather heavy particles with masses by an order
of magnitude larger than the  inflaton mass can be produced quite
copiously
(see also \cite{KRT98}).
Applying these results to the case of our interest, we find
that the stable very heavy particles, $m_\phi \alt m_X \alt 10\, m_\phi$,
will be generally produced in excess in this process.

However, if the parametric resonance is ineffective for some reason,
and one
estimates the particle number density after inflation at the level
of initial
conditions used in Refs. \cite{KT}, one finds that $\Omega_X$
might prove to be of the acceptable magnitude. This level is saturated
by the fundamental process of particle creation during inflation from
vacuum fluctuations and it is the same process which generated
primordial large scale density perturbations.
Parametric resonance for $X$ particles is turned off if $X$ field is
either
a fermion field or its coupling to inflaton is small,
$g^2 \ll 10^4 (m_X/m_\phi)^4 (m_\phi/M_{\rm Pl})^2 $ \cite{KT}.
(Parametric creation of fermions was considered in Refs. \cite{GK98}.
While it is not a ``resonance'' process, it might be useful for creation
of very heavy fermions.)

Particle creation from vacuum fluctuations during inflation
(or in the de Sitter
space) was extensively studied \cite{pc_dS1,pc_dS2}, usually in the case
of small $m_X$ and in application to  generation of density fluctuations
necessary for the large scale structure formation. The
characteristic quantity which is usually cited, the variance of
the field, $\langle X^2 \rangle$, is defined by an expression similar to
Eq. (\ref{nX}). In the typical case the
difference is given by the
factor $2\sin^2(\omega_k\eta)/\omega_k$ in the integrand.
For example, for the scalar Bose field with the minimal coupling to the
curvature,
$\langle X^2 \rangle = 3H_i^4/8\pi^2m_X^2$ if
$m_X \ll H_i$\cite{pc_dS1,pc_dS2}.  For massless self-interacting field
$\langle X^2 \rangle \approx 0.132 H_i^2/\sqrt{\lambda}$ \cite{SY}.
Particle creation in the settings close to our problem, but
for the specific case of the Hubble dependent effective
mass, $m_X(t) \propto H(t)$, was considered in Ref. \cite{LR}.

In general,
if $m_X \sim H_i \approx m_\phi$, we can again expect \cite{KT98}
that the particle density will be given by Eq. (\ref{n_fr}).
In general, the coefficient $C$ in this equation will be
a function of the ratio $H_i/m_X$, of the coupling constant $\xi$,
and will depend on details of the
transition between inflationary and matter (or radiation) dominated
phases, etc. For $m_X > H_i$ an exponential suppression is expected.
Detailed numerical study of X-particle abundance created from
the vacuum in inflationary
cosmology is the subject of this Section of the paper.

The case of large $m_X$ with application to the dark matter problem was
already considered in Ref. \cite{CKR}. However, the analysis
was restricted to the case of conformally coupled scalar particles only,
and approximations were made to the extent that
the fixed de Sitter background was matched to the subsequent
radiation or matter dominated expansion either as an instantaneous
transition or with the help of some smoothing function.
Particle creation in the case of large $m_X$ can be especially
sensitive to (some extent arbitrary) details of such junction procedure.

In the present paper, considering particle creation as described
by Eqs. (\ref{nX})-(\ref{beta_f}), we do not
make any approximations. We assume a specific model of inflation,
namely ``chaotic'' inflation \cite{al83}, and find numerically the
exact evolution of the scale factor in this model.
We consider both  the ``massive'' inflaton with the scalar potential
$V(\phi) = m^2\phi^2/2$, and the ``massless'' inflaton with the potential
$V(\phi) = \lambda\phi^4/4$. The  normalization to the observable
large scale
structure requires $m^2/M_{\rm Pl}^2 \approx 10^{-12}$ in the former model
and $\lambda \approx 10^{-13}$ in the latter model. We consider
creation of conformally and minimally coupled scalar X-particles, as
well as
creation of X-fermions.

\begin{figure}
\leavevmode\epsfysize=5.5cm \epsfbox{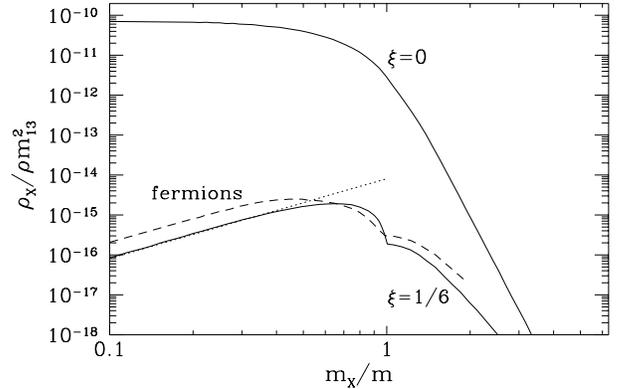}
\caption{Ratio of the energy density in $X$-particles to the total
energy density at late times in a model with the massive inflaton,
$V(\phi) = m^2\phi^2/2$, as a function of X particle mass, $m_X$.
For the inflaton mass we defined $m_{13} \equiv m/10^{13}$ GeV.
The dotted line is the low mass asymptotic, Eq. (\ref{n_fr}).
}
\label{fig:m_infl}
\end{figure}

\begin{figure}
\leavevmode\epsfysize=5.5cm \epsfbox{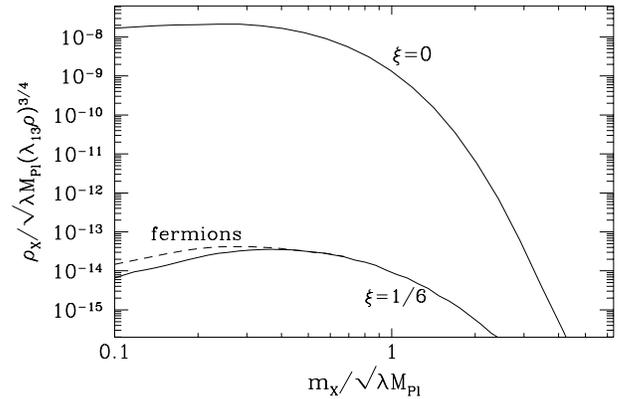}
\caption{Energy density of created $X$-particles in the
massless inflaton model,
$V(\phi) = \lambda\phi^4/4$. We defined $\lambda_{13} \equiv
\lambda/10^{-13}$.
}
\label{fig:l_infl}
\end{figure}

Our results are summarized in Fig.~\ref{fig:m_infl} (massive inflaton)
and Fig.~\ref{fig:l_infl} (massless inflaton).
We define the energy density in X-particles
at late times as $\rho_X = m_X n_X$.
In the case of the massive inflaton we normalized $\rho_X$ by the total
energy density. The ratio of those two quantities reaches some
asymptotic value and then
remains constant in a matter dominated universe,
and can be measured at any sufficiently late time in the numerical
simulation
which does not include the inflaton decay. For each value of $m_X$
we integrated equations (\ref{nX})-(\ref{beta_f}) until the asymptotic
constant value of this ratio was reached.
Similarly, the convenient quantity
to be measured in the massless inflaton model is proportional to
$\rho_X/\rho^{3/4}$,
which also becomes constant at late times.
The curves labeled by $\xi=0$
and $\xi=1/6$ correspond to the minimal and the conformal coupling
to gravity,
respectively. The dashed line describes creation of fermions.

We see that particle production is most effective in the case of
scalar particles with minimal coupling, exceeding $\rho_X$ generated in
other models by many orders of magnitude if $m_X \alt m$.
Note also that the ratio of the energy
density in $X$-particles to the total energy density is
independent upon $m_X$ at $m_X < m$ in this model. This might seem
as easily
understandable. Indeed, magnitude of  fluctuations generated during
inflation in de Sitter space is $\langle X^2 \rangle \propto m_X^{-2}$ if
$m_X \ll H_i$. Multiplying this by $m_X^{2}$ one would find
$\rho_X \approx m_X^2 \langle X^2 \rangle =$ const,
seemingly in agreement
with observed behavior, Fig.~\ref{fig:m_infl}. However,this
agreement is rather coincidental since it is possible to relate $\rho$
and $\langle X^2 \rangle$ in this simple manner at late
times only.

In our numerical calculations we had found
that the variance $\langle X^2 \rangle$
of the field $X$  measured at the end of inflation is independent upon
$m_X$ if the mass of $X$ is small.
At some later epoch when $H \approx m_X$, which will be long after
the end of
inflation if X is a light field,
the field $X$ starts to oscillate on all scales, including $k=0$.
Only at this time, which we denote by  $t_X$, all field fluctuations are
transformed into non-zero particle density and we can use
$\rho_X = m_X n_X \approx m_X^2 \langle X^2 \rangle $.
The variance of $X$ fluctuations
was unchanged on large scales, starting from the end of inflation
down to the
time $t_X$. So, when the field starts to oscillate,
$\rho_X \propto m_X^2$. However, the  energy density of the inflaton
field,
$\rho = 3H^2/8\pi G$ decreased during this time interval in proportion
to $H^2(t_X)/H^2(0) \approx m^2_X/H^2(0)$. That is why the ratio of
the energy
density in $X$-particles to the total energy density does not depends
on $m_X$ when measured at $t > t_X$.

\begin{figure}
\leavevmode\epsfysize=5.5cm \epsfbox{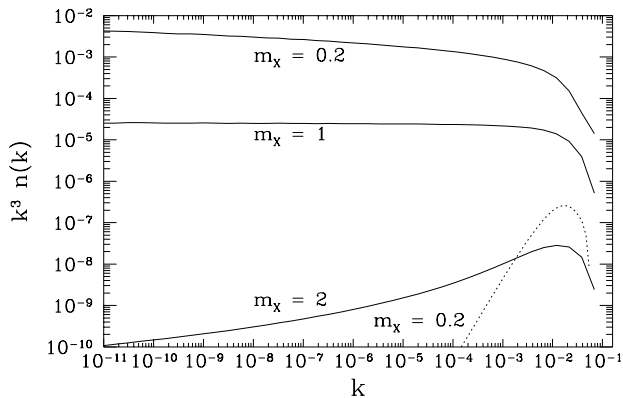}
\caption{Spectrum of created particles, $k^3n(k)$, in a model with massive
inflaton is shown for several choices of the mass of scalar
X-particle with
the minimal coupling (solid lines) and the conformal coupling
(dotted line).
Masses and momenta, $k$, are given in units of the inflaton mass.
}
\label{fig:nk}
\end{figure}

Variance of the  field X is different from the usually
calculated for the fixed de Sitter inflationary background because
we consider the actual evolution of the scale factor and the value of
the Hubble parameter is not constant
during inflation, being larger at earlier times. Correspondingly,
the number
of created particles per decade of $k$ grows logarithmically towards
small k if $m_X$ is small. (Power spectrum behaves similarly.)
For definiteness, we restricted our calculations to $\sim 10^{15}$ for
the expansion factor during inflation. The  examples of the particle
number,
$k^3 n_X (k)$ for several values of $m_X$ are shown in Fig.~\ref{fig:nk}
at the moment corresponding to 10 completed inflaton oscillations.
The particle momentum is measured in units of the inflaton mass.
In contrast to this, in fixed de Sitter background
$4\pi^2\langle X^2 \rangle \approx H^2 \int d \ln k \,(k/H)^{3-2\nu}$ with
$3-2\nu \approx 2m_X^2/3H^2$ at small $m_X$ and consequently
$\langle X^2 \rangle \propto 1/m_X^2$.
When our code was tested on fixed de Sitter background it reproduced
the proper power spectrum exactly. Note that the power spectrum in
fixed de Sitter background grows towards large values of $k$,
which is opposite to the behavior of Fig.~\ref{fig:nk}.

Therefore, calculations which would be based on customary 
procedure of matching a fixed
de Sitter background to a subsequent Friedmann stage would give wrong
results in this case. Note, however, that the result depends upon duration
of inflationary stage, so our results in the case of scalars with minimal
coupling  and $m_X < m$ should be considered only as a lower bound on
X-particle abundance. 

Matching is also dangerous in the case of large
$m_X$, and its influence was studied in detail in Ref. \cite{CKR}.
When the change is too abrupt, it generates artificial particles.
This can easily happen for $m_X > m$, see e.g. \cite{wrong_paper}
where excessive production was found.
At $m_X \agt m$ the number of created particles decreases exponentially
with $m_X$. For the case of massive inflaton our result shown
in Fig.~\ref{fig:m_infl} can be fitted as
\begin{equation}
\frac{\rho_X }{\rho} \approx 10^{-10} \frac{m_X}{m} m_{13}^2
\exp (-2\pi m_X/m) \, \, ,
\label{exp}
\end{equation}
where $m_{13} \equiv m/10^{13}$ GeV. (Note that the fitting formulae
is approximate and arbitrary to some extent).

As Fig.~\ref{fig:nk} shows, the power spectrum of fluctuations
in X-particles is almost scale independent at small k if
$m_X/m \approx 1$. Therefore, if such particles constitute a considerable
fraction of dark matter, this fluctuations will be transformed into
isocurvature density perturbations at late times and can affect 
large scale structure formation. Isocurvature
fluctuations produce 6 times larger angular temperature fluctuations
in cosmic microwave background radiation (CMBR) for the same amplitude 
of long-wavelength density perturbations 
compared to the adiabatic case \cite{SS84}.
Therefore, to fit observations by a single spectrum,  
the mass fluctuation spectrum in isocurvature cold dark matter cosmology 
must be tilted (with respect to scale invariant spectrum) to favor
smaller scales.  

Let the power spectrum of the field
fluctuations be $k^3P_X(k) \propto k^\beta$. To fit to the second 
moments of the large-scale mass and (not contradict to) cosmic microwave
background distributions requires $\beta \agt 0.25$. Models
with $\beta $ ranging from 0.3 to 0.6 were constructed and 
considered in  Refs. \cite{ML,Peebles}. 
It is interesting that in our case
the spectrum is indeed correctly tilted, see Fig.~\ref{fig:nk},
for $m_X$ which is somewhat larger than the inflaton mass,
On the smallest scale of our integration, $k \approx 10^{-15}$
we find $\beta \approx 0.1$ for $m_X/m =2$ and $\beta \approx 0.3$ 
for $m_X/m =3$. Since $\beta$ decreases with the length scale increasing,
on scales corresponding to the current horizon $\beta$ will be smaller.
Because of that the amplitude of perturbations will be also too small
on cosmological scales when $\beta$ will be reaching desired magnitude,
$\beta \agt 0.25$. Indeed,
density fluctuations, $\delta \rho_X/\rho_X$, which are roughly
proportional to spectra shown in Fig.~\ref{fig:nk}, are reaching 
unity near the break
in the power spectrum ($k \sim 1$ at the end of inflation) and
on the scales of interest amplitude of fluctuations became 
too small already at $m_X/m =  3$.
However, their magnitude may be just right, 
$\delta \rho_X/\rho_X \approx 10^{-5}$, on scales $k \agt 10^{-25}$
for the case $m_X/m \approx 2$ (or slightly larger) if $\Omega_X \approx 1$, 
see Fig.~\ref{fig:nk}.  Magnitude
can not correspond to the observable on the galaxy scales since then
there will be too much fluctuations in CMBR. This gives a constraint on the
models. X-particles with minimal coupling to gravity, $m_X > 2m$
and $\Omega_X \approx 1$ are excluded. (Such particles with small
contribution to $\Omega$ may exist though.) However, the magnitude
of the density fluctuations induced in the process of  X particle 
creation can correspond to the observable on the horizon scale, and be 
responsible for fluctuations in CMBR.  X-particles with 
$m_X/m \approx 2$ and $\Omega_X \approx 1$ can give such 
contribution naturally. (Source for the galaxies seeds
has to be different in this case.) 
Note that the density autocorrelation function 
is proportional to the square of 
the field autocorrelation function in the present case, and the density 
fluctuations are non-Gaussian.

In the case of fermions or scalar particles with conformal coupling, the
number density of created particles
has a kink at $m_X =m$ in the model with massive inflaton,
see Fig.~\ref{fig:m_infl}.
This signifies that different mechanisms are responsible for particle
creation at $m_X > m$ and  $m_X < m$. Particles heavier than the inflaton
are created during inflation, while particles lighter than the inflaton
are created
after inflation during the regular Friedmann stage of expansion. We see
that particle creation during this latter stage is actually more
effective
($n_X$ rapidly increases when $m_X$ drops slightly below $m$).
Hence, we can assume that the number density of created particles at
$m_X \ll m$
is given by Eq. (\ref{n_fr}) with $C_\alpha = C_{2/3}$.
($C_{2/3} \approx 9.7\times 10^{-4}$ for the conformal scalars,
see Fig.~\ref{fig:C}.)
Since the total energy
density is given by $\rho = 3 H^2/8\pi G$, we find for $\alpha =2/3$
\begin{equation}
\frac{\rho_X}{\rho} = \frac{8\pi}{3}\,\frac{ m_X^2}{M_{\rm Pl}^2}\,
C_{3/2}
\, .
\label{lma}
\end{equation}
This function is plotted in Fig.~\ref{fig:m_infl} by the dotted line
which shows that already at $m_X < 0.5\, m$ the number density of created
particles is given by
 Eq. (\ref{n_fr}), indeed. Note that in Ref. \cite{CKR}
all particle creation was attributed to the transition from the
inflationary
to matter dominated phase, while we find that for $m_X < m$
the dominant particle creation occurs
at a later stage, when $H \approx k \approx m_X$ (for $m_X < m$).
The  correct
interpretation allows us to identify the low mass asymptotic,
Eq. (\ref{lma}), see also \cite{KT98}.

Scalar particles with minimal coupling to gravity are produced
predominantly
during the inflationary stage for all $m_X$, and the resulting number
density is much higher, see Fig.~\ref{fig:m_infl}.

\begin{figure}
\leavevmode\epsfysize=5.5cm \epsfbox{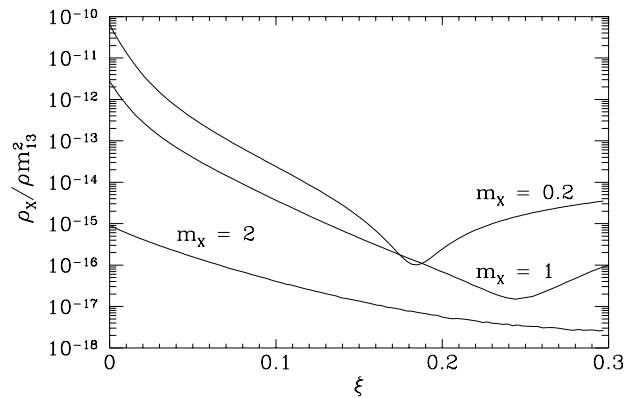}
\caption{Ratio of the energy density in $X$-particles to the total
energy density at late times in a model with the massive inflaton,
$V(\phi) = m^2\phi^2/2$, as a function of $\xi$.}
\label{fig:xi_scan}
\end{figure}

Conformal coupling, $\xi =1/6$, is a stable point of renormalization
group equations. Otherwise $\xi$ is a running coupling constant
and there is no reason to expect the coupling to be, say, minimal
at mass scales of inflation. The
dependence of X-particle abundance on the coupling to the curvature scalar
is shown in Fig.~\ref{fig:xi_scan}.

The situation in the case of the massless inflaton,
$V(\phi )=\lambda \phi^4/4$,
is  quite similar. However, now there is no distinctive mass scale
in the equations of motion for the scale factor, and the kink
in the function $n_X (m_X)$ does not appear. The quantity
$\sqrt{\lambda} \phi$
might be considered as an effective inflaton mass, but it changes
during the
evolution. For definiteness we plot $n_X$ as a function of
$m_X/\sqrt{\lambda}M_{\rm Pl}$. Note that the transition from the
inflationary to the
radiation dominated stage occurs at $\phi \approx 0.35 M_{\rm Pl}$
\cite{KT} and this may define the more natural mass scale for the ratio
$m_X/m_{\rm eff}(\phi )$. Indeed, we observe that $n_X$ reaches its
maximum
when $m_X \approx 0.35 \sqrt{\lambda}M_{\rm Pl}$, see
Fig.~\ref{fig:l_infl}.

We find that the number density of produced fermions in both cases
is approximately equal to the number density of scalar
particles with conformal coupling. The main difference amounts to a spin
factor of two.

\section{Present density of X-particles}

 Let us estimate the present day number density of X-particles.
Consideration is somewhat different for massive and massless inflaton
models.

{\it 1. Massive inflaton, $V(\phi) = m_\phi^2 \phi^2/2 $.}

In this case the inflation
is followed by the stage of matter domination which lasts until inflaton
decays and thermal equilibrium is established. Let us denote the
temperature
at the beginning of radiation dominated stage as $T_*$.
Actually, a transition to the radiation dominated universe does not
necessarily occur with the final state being in thermal equilibrium.
Complications arise if there are light bosons
in a theory, $m_B \ll m_\phi$, even relatively weakly coupled to the
inflaton,
$g^2 \agt 10^4 m_\phi^2/M_{\rm Pl}^2 \sim 10^{-8}$. In this case
the inflaton will decay via parametric resonance
after just a several oscillations and relativistic degrees of freedom
start to dominate the equation of state.
This happens typically when the energy
density in the inflaton oscillations is red-shifted by a factor
$r \approx 10^{-6}$ compared
to the value $m_\phi^2 M_{\rm Pl}^2$ \cite{KLS,KT}. Matter is still far
from being in thermal equilibrium, but since Universe expansion
already at this point changes to radiation dominated it is convenient
to characterize
this radiation dominated stage by an equivalent ``temperature'',
$T_* \sim r^{1/4} \sqrt{m_\phi M_{\rm Pl}}$.
In any case, $T_*$ is the model dependent parameter and we leave it as
such.

Until the moment when the equation of state changes from matter
dominated to radiation dominated, the ratio of the energy density
in X-particles to the total energy density retains its value reached
at the
end of inflation, $\rho_X/\rho = $ const. This ratio is plotted in
Fig.~\ref{fig:m_infl} for different models as a function of model
parameters
and it defines the present day ratio of $\rho_X$ to the critical density
in the Universe, $\rho_c$, as follows:
\begin{equation}
\Omega_X h^2 = \frac{\rho_X}{\rho} \frac{T_*}{\gamma T_0} \Omega_R
h^2 \, ,
\label{rh1}
\end{equation}
where $T_0$ is present temperature of cosmic microwave background, and
$\Omega_R h^2 = \rho_{\rm rad} h^2 /\rho_c \approx 4.31 \times 10^{-5} $
is the fraction of critical density that is in radiation today
\cite{KT_book}. The
following definitions are used here $\rho_c = 3H_0^2 M^2_{\rm Pl}/8\pi$
and $H_0 = 100 h$ km sec$^{-1}$ Mpc$^{-1}$.
Parameter $\gamma$ describes the change in entropy per comoving volume
for the whole cosmological history starting from $T=T_*$ down to the
present
day, i.e.  $s_0 \equiv \gamma s_*$. In simple cosmological models,
$\gamma \sim 1$. However, in models with subsequent latent heat release
(due to phase transitions or due to decay of long living particles,
e.g. moduli fields) $\gamma$ can be very large.

Let us assume for illustration purposes that $T_* = 10^9$ GeV, i.e.
$(T_*/T_0)(\Omega_R h^2) \approx 1.8 \times 10^{17}$.
(This is the highest allowed reheating temperature when gravitinos in
supergravity theories are not overproduced.)
With $\gamma \sim 1$ we find that $\Omega_X h^2 \sim 1$ if
$m_X/m \approx 2$ in models with scalar X-particles with conformal
coupling,
or if X-particles are fermions, see Fig.~\ref{fig:m_infl}.
For minimally coupled scalars this value of $\Omega_X$ is reached for
$m_X/m \approx 3$.

To summarize this subsection:
$\Omega_X h^2$ is found multiplying the data of Fig.~\ref{fig:m_infl}
by a factor
$1.8 \times 10^{17} (T_* / \gamma 10^9 \, {\rm GeV}) m^2_{13}$.

{\it 1. Massless inflaton, $V(\phi) = \lambda \phi^4/4 $.}

Before and after the decay of the inflaton the Universe expands as
radiation
dominated. In Section III for the state which precedes the decay of
inflaton
oscillations we had calculated the quantity
$\rho_X/\sqrt{\lambda} M_{\rm Pl} \rho^{3/4}$, where $\rho$ is the
inflaton
energy density. After the decay of the inflaton, in a state of
thermal equilibrium we can use
\begin{equation}
s_* = \frac{4}{3} \left(\frac{\pi^2 g_*}{30}\right)^{1/4} \,
\rho_*^{3/4} \, ,
\label{rh2}
\end{equation}
where $g_*$ is the number of relativistic degrees of freedom at
temperature
$T_*$. Assuming instantaneous transition between those states,
$\rho_* = \rho$, and again using
$\Omega_X = (\rho_X/T_0s_0)(T_0s_0/\rho_c)$ together with
$s_0 \equiv \gamma s_*$, we find
\begin{equation}
\Omega_X h^2 = \frac{\sqrt{\lambda} M_{\rm Pl}}{\gamma T_0}
\left(\frac{\pi^2 g_*}{30}\right)^{-1/4}
\frac{\rho_X}{\sqrt{\lambda} M_{\rm Pl} \rho^{3/4}}
\Omega_R h^2 \, .
\label{rh3}
\end{equation}
We see that $\Omega_X h^2$ can be found by
multiplying the data plotted in Fig.~\ref{fig:l_infl} by a factor
$2.5 \times 10^{20} \lambda^{5/4}_{13} /\gamma g_{200}^{1/4}$,
where we defined $g_{200} \equiv g_*/200$, and used
$\sqrt{\lambda} M_{\rm Pl}/ T_0 \approx 1.6 \times 10^{25}
\sqrt{\lambda_{13}}$. Note that
$\sqrt{\lambda} M_{\rm Pl} \approx 3.9 \times 10^{12} \sqrt{\lambda}_{13}$
GeV, so that already with $\gamma =1$ we shall have
$\Omega_X < 1$ if $m_X \agt 4 \times 10^{13}$ GeV.

\section{X-particles and UHE cosmic rays.}

Why were the Ultra High Energy cosmic rays observed above the
Greisen-Zatsepin-Kuzmin energy cut-off? One
solution to the problem might be provided, for example,
by the existence of some
exotic particles which are able to propagate (evading the GZK bound) from
cosmological distances and yet interact in the Earth's
atmosphere like a hadron. A particle with such conflicting properties
was found
in a class of supersymmetric theories \cite{farrar}. Alternatively,
high energy cosmic rays may have been produced locally within the GZK
distance. One possibility
is connected with the destruction of topological defects
\cite{td}, while another one is connected with decays of primordial
heavy long-living particles \cite{KR,BKV}. The candidate $X$-particle must
obviously  obey constraints on mass, density and lifetime.

In order to produce cosmic rays in the energy range
$E \agt 10^{11}$ GeV, the mass of
$X$-particles has to be very large, $m_{X} \agt 10^{13}$~GeV
\cite{KR,BKV}.
The lifetime, $\tau_{X}$, cannot be much smaller than
the age of the Universe, $\tau \approx 10^{10}$~yr. With such a short
lifetime, the observed flux of UHE cosmic rays will be
generated with the rather low density of $X$-particles,
$\Omega_{X} \sim 10^{-12}$,
where $\Omega_{X} \equiv m_{X} n_{X}/\rho_{\rm crit}$, $n_X$ is the
number density
of X-particles and $\rho_{\rm crit}$ is the critical density.
On the other hand, X-particles must not overclose the Universe,
$\Omega_{X} \alt 1$.
With $\Omega_{X} \sim 1$, the X-particles may play the role of cold dark
matter and the observed flux of UHE
cosmic rays can be matched if $\tau_{X} \sim 10^{22}$~yr.
The allowed windows are quite wide \cite{KR}, but on the exotic
side, which
may give rise to some problems.

The possibility that X-particles were generated from the vacuum by
gravitational interactions \cite{KT98,CKR} may answer to the question why
their abundance should lie in the interesting range $\Omega_{X} \alt 1$
and was considered in detail in the present paper.

The problem of the particle physics mechanism responsible for a long but
finite lifetime of very heavy particles can be solved in several ways.
For example, otherwise conserved quantum number carried by  X-particles
may be broken very weakly
due to instanton transitions \cite{KR}, or quantum gravity (wormhole)
effects \cite{BKV}. If instantons are responsible for $X$-particle
decays, the lifetime is estimated as
$\tau_{X} \sim m_{X}^{-1}\cdot \mbox{exp}(4\pi/{\alpha_{X}})$,
where $\alpha_{X}$ is the coupling constant of the relevant gauge
interaction.
The lifetime will fit the allowed window if the coupling constant
(at the scale $m_{X}$) is $\alpha_{X} \approx 0.1$ \cite{KR}.

A class of natural candidates for superheavy long-living particles
which arise in string and M theory was re-evaluated recently in Ref.
\cite{BEN98} and particles with desired mass and long life-time were
identified. Another interesting candidates were found among adjoint
messengers in gauge mediated supergravity models \cite{CKR,HYZ98}.

\section{Conclusions}

We have shown that the very weakly interacting superheavy X-particles
with $m_X = ({\rm a~few}) \cdot 10^{13}$ GeV may naturally constitute a
considerable fraction of Cold Dark Matter. These particles are
produced in the early Universe from vacuum fluctuations
during or after inflation. Related density fluctuations 
may have left an imprint in fluctuations of cosmic microwave background 
radiation if scalar X-particles with minimal coupling to gravity
are approximately twice heavier than the 
inflaton and $\Omega_X \sim 1$.
Decays of X-particles may explain UHE cosmic rays phenomenon. 

Our hypothesis has unique observational consequences. If UHE cosmic rays
are indeed due to the decay of these superheavy particles,
there has to be a new sharp cut-off in
the cosmic ray spectrum at energies somewhat smaller than $m_X$. Since
the number
density $n_X$ depends exponentially upon $m_X/m_\phi$, the position
of this
cut-off is fixed  and can be predicted to be near
$m_\phi \approx 10^{13}$ GeV, the very shape of the cosmic
ray spectrum beyond the GZK cut-off being of quite generic form
following from the QCD quark/gluon fragmentation. Next generation 
experiments, like the Pierre Auger Project \cite{auger}, High Resolution
Fly's Eye \cite{FlysEye}, the Japanese Telescope
Array Project \cite{JapArray},  may prove to be able to discover this
fundamental phenomenon.

We conclude that the observations of Ultra High Energy cosmic rays
can probe
the spectrum of elementary particles in the superheavy range and can give
an unique opportunity for investigation of the earliest epoch of
evolution of the Universe,
starting with the amplification of vacuum fluctuations during inflation
through fine details of gravitational interaction and down to the physics
of reheating.


We are grateful to S. Khlebnikov, A. Linde and A. Riotto for 
useful discussions.
V.~A.~Kuzmin and I.~I.~Tkachev thank Theory Division at CERN for
hospitality where this work was initiated.
The work of V.~K. was supported in part by
Russian Foundation for Basic Research under  Grant 95-02-04911a.
I.~I.~T.  was supported in part by the U.S. Department of Energy
under Grant DE-FG02-91ER40681 (Task B) and by the National Science
Foundation under Grant PHY-9501458.

\end{document}